\documentclass[
    ,final            
  ]
  {aipproc}

\layoutstyle{6x9}

\usepackage{amsmath}
\usepackage{bm}

\def\centeron#1#2{{\setbox0=\hbox{#1}\setbox1=\hbox{#2}\ifdim
\wd1>\wd0\kern.5\wd1\kern-.5\wd0\fi
\copy0\kern-.5\wd0\kern-.5\wd1\copy1\ifdim\wd0>\wd1
\kern.5\wd0\kern-.5\wd1\fi}}
\def\lsim{\;\centeron{\raise.35ex\hbox{$<$}}{\lower.65ex\hbox{$\sim$}}\;}
\def\gsim{\;\centeron{\raise.35ex\hbox{$>$}}{\lower.65ex\hbox{$\sim$}}\;}

\begin{document}

\title{Relativistic Corrections to $\bm e^{\bm +}\bm e^{\bm -}\bm\to \bm
J\bm/\bm\psi \bm +\bm \eta_{\bm c}$ in a Potential
Model\protect\footnote{Talk presented by
G.~T.~Bodwin.}
}



\author{Geoffrey T.~Bodwin\protect\footnote{Work in the High Energy
Physics Division at Argonne National Laboratory is supported by the
U.~S.~Department of Energy, Division of High Energy Physics, under
Contract No.~W-31-109-ENG-38.} }{   address={High Energy Physics
Division, Argonne National Laboratory, 9700 South Cass Avenue, Argonne,
Illinois 60517}
}

\author{Daekyoung Kang}{
  address={Department of Physics, Korea University, Seoul 136-701, Korea}
}

\author{Taewon Kim}{
  address={Department of Physics, Korea University, Seoul 136-701, Korea}
}

\author{Jungil Lee}{
  address={Department of Physics, Korea University, Seoul 136-701, Korea}
}

\author{Chaehyun Yu}{
  address={Department of Physics, Korea University, Seoul 136-701, Korea}
}

\begin{abstract}
We compute relativistic corrections to the process $e^+e^-\to J/\psi 
+\eta_c$ and find that they resolve the discrepancy between theory and 
experiment.
\end{abstract}

\maketitle


The disagreement between theory and experiment for the exclusive
double-charmonium process $e^+e^-\to J/\psi+\eta_c$ has, for a number of
years, been one of the largest discrepancies in the standard model. The
production cross section times the branching fraction into two or more
charged tracks $\sigma(e^+e^-\to J/\psi+\eta_c)\times B_{>2}$, has been
measured by the Belle Collaboration to be $25.6\pm 2.8\pm 3.4~\hbox{fb}$
(Ref.~\cite{Abe:2004ww}) and by the {\it BABAR} Collaboration to be
$17.6\pm 2.8^{+1.5}_{-2.1}~\hbox{fb}$ (Ref.~\cite{Aubert:2005tj}). In
contrast, nonrelativistic QCD (NRQCD) factorization \cite{BBL}
calculations at leading order in $\alpha_s$ predict cross sections of
$3.78\pm 1.26~\hbox{fb}$ (Ref.~\cite{Braaten:2002fi}) and $5.5~\hbox{fb}$
(Ref.~\cite{Liu:2002wq}). The differences between these calculations
arise from different choices of $m_c$, NRQCD matrix elements, and
$\alpha_s$ and from the fact that the calculation of
Ref.~\cite{Braaten:2002fi} includes QED effects, while that of
Ref.~\cite{Liu:2002wq} does not. An important recent development is the
calculation of the corrections of next-to-leading order in
$\alpha_s$ (Ref.~\cite{Zhang:2005ch}), which yield a $K$~factor of about
$1.96$. However, even if one includes this $K$~factor, a significant
discrepancy remains.

It is known from the work of Ref.~\cite{Braaten:2002fi} that relativistic
corrections to the process are potentially large. The first relativistic
correction appears at order $v^2$, where $v$ is the heavy quark or
antiquark velocity in the quarkonium rest frame. ($v^2\approx 0.3$ for
charmonium.) In Ref.~\cite{Braaten:2002fi}, the order-$v^2$ corrections
are estimated to give a $K$~factor $2.0_{-1.1}^{+2.9}$. The large
uncertainties arise because of large uncertainties in the relevant
nonperturbative NRQCD matrix element of relative
order $v^2$. If the order-$v^2$ $K$~factor is indeed large, then this
casts doubt on the convergence of the $v$ expansion of NRQCD. In what
follows, we address the uncertainty in the order-$v^2$ matrix through a 
potential-model calculation and the convergence of the $v$ expansion 
through resummation.

The NRQCD matrix element of leading-order in $v$ for production (or 
decay) of the $\eta_c$ is related to the wave function at the origin 
$\psi(0)$:
\begin{equation}
\psi(0)\equiv                             
\int\frac{d^3p}{(2\pi)^3}                                   
\widetilde{\psi}(\bm{p})                               
=\frac{1}{\sqrt{2N_c}}
\langle 0|\chi^\dagger\psi|\eta_c\rangle.
\end{equation}
Here $\psi$ annihilates a heavy quark, $\chi^\dagger$ annihilates a
heavy antiquark, and $\psi({\bf x})$ and $\tilde\psi({\bf p})$ are
coordinate-space and momentum-space Schr\"odinger wave functions,
respectively. A similar matrix element appears in NRQCD expressions for
$J/\psi$ production and decay. The corresponding NRQCD matrix elements 
of higher order in $v^2$ are given by
\begin{equation}
\psi^{(2n)}(0)\equiv
\int\frac{d^3p}{(2\pi)^3}
\,\bm{p}^{2n}
\widetilde{\psi}(\bm{p})=
\frac{1}{\sqrt{2N_c}}
\langle 0|\chi^\dagger(-\bm{\nabla}^2)^n\psi|\eta_c\rangle.
\end{equation}
We use the following notation for such matrix elements: $ \langle
\bm{p}^{2n}\rangle\equiv \psi^{(2n)}(0)/\psi(0)$ and $\langle
v^2\rangle=\langle \bm{p}^2\rangle/m_c^2$. Previous attempts to
determine $\psi^{(2)}(0)$ from phenomenology, from lattice measurements,
and from the Gremm-Kapustin relation \cite{Gremm:1997dq} have resulted
in large uncertainties. Even the sign of $\psi^{(2)}(0)$ is not known with
great confidence from these methods.

Our strategy is to use a potential-model calculation, with the Cornell
potential \cite{Eichten:1978tg}, to determine $\psi^{(2)}(0)$. Details
of this calculation can be found in Ref.~\cite{Bodwin:2006dn}. If the
model potential is the exact static $Q\bar Q$ potential, then the errors
in the potential model are of relative order $v^2$
(Ref.~\cite{Brambilla:1999xf}). With an appropriate choice of parameters,
the Cornell potential provides a good fit to lattice data for the static
$Q\bar Q$ potential.

The matrix element $\psi^{(2)}(0)$ contains a linear UV divergence and
must be regulated. Because existing calculations in NRQCD of order
$\alpha_s$ and higher make use of dimensional regularization, we
ultimately want to obtain a dimensionally regulated matrix element. We
introduce two methods to achieve this.

Method~1 requires knowledge only of the wave function. In this method,
we first regulate using a simple, analytic momentum-space hard cutoff
$\Lambda^2/(\bm{p}^2+\Lambda^2)$. Then, we calculate the difference
between $\psi^{(2)}(0)$ in hard-cutoff and dimensional regularization,
which we call $\Delta\psi^{(2)}(0)$. Since $\Delta\psi^{(2)}(0)$ gives
the difference between UV regulators, it is dominated by large momenta
and can be computed in perturbation theory. We subtract
$\Delta\psi^{(2)}(0)$ from the hard-cutoff result. In order to obtain
the dimensionally regulated result, we must extrapolate this subtracted
expression to $\Lambda=\infty$ because our computation of
$\Delta\psi^{(2)}(0)$ does not include contributions that are suppressed
as $1/\Lambda$.

Method~2 is applicable only to potential models. It requires knowledge of
the binding energy and the potential, in addition to the wave function.
In this approach, we use the Bethe-Salpeter equation to expose an 
explicit loop from the wave function in the expression for the matrix 
element. Then, we regulate that loop dimensionally. The result is 
equal to that which would be obtained from the Gremm-Kapustin relation, 
but for the binding energy of the potential model. 

The results from Method~1 and Method~2 agree well numerically and yield 
$\psi^{(2)}(0)=0.118\pm 0.024\pm 0.035~\hbox{GeV}^{7/2}$ and $\langle 
\bm{p}^2\rangle = 0.50\pm 0.09\pm 0.15~\hbox{GeV}^2$, which imply for 
$m_c=1.4~\hbox{GeV}$ that $\langle v^2\rangle\approx 0.25\pm 0.05\pm 
0.08$. This last result is in good agreement with expectations from the 
NRQCD $v$-scaling rules. In all of these results, the first error bar 
reflects the uncertainty in the input potential-model
parameters and the wave function at the origin, and the 
second error bar reflects relative-order-$v^2$ corrections that 
have been neglected. This is the first determination of
$\psi^{(2)}(0)$ with small enough uncertainties to be useful 
phenomenologically.

We can extend Method~2 to matrix elements of higher order in $v$. By 
using the equation of motion, dimensional regularization, and the
scalelessness of the individual terms in the Cornell potential, we obtain 
the simple relation
\begin{equation}
\langle \bm{p}^{2n}\rangle =
\langle \bm{p}^2\rangle^n.
\end{equation}
This relation allows one to resum a class of the relativistic corrections
to $S$-wave quarkonium decay and production amplitudes to all orders in
$v$.

We now apply these results to the relativistic corrections to
$\sigma[e^+e^-\to J/\psi+\eta_c]$. These corrections  arise in two ways.
First, they appear directly in the process $e^+e^-\to J/\psi +\eta_c$
itself. Second, they enter indirectly through $|\psi(0)|^2$, which
appears as a factor in the contribution of leading order in $v^2$.  The
quantity $|\psi(0)|^2$ is determined from the experimental value for the
width for $J/\psi \to e^+e^-$ and the theoretical expression for that
process, which is affected by relativistic corrections.

Our preliminary results, including the effects of QED, as well as QCD,
contributions are as follows. The resummed relativistic corrections to
$\sigma[e^+e^-\to J/\psi +\eta_c]$ itself yield a $K$~factor $1.34$,
while the resummed relativistic corrections to $|\psi(0)|^2$ yield a 
$K$~factor $1.32$. In both cases, the resummation of contributions beyond
order $v^2$ has only about a $10\%$ effect on the $K$~factors, which
indicates that $v$ expansion of NRQCD converges well for these
processes. Combining the resummed relativistic-correction $K$~factors
with the relative-order-$\alpha_s$ $K$~factor $1.96$, we obtain a complete 
$K$~factor $4.15$. Applying this to the calculation of
Ref.~\cite{Braaten:2002fi} and using the most recent value for
$\Gamma[J/\psi\to e^+e^-]$ (Ref.~\cite{Yao:2006px}), we obtain a
prediction $\sigma[e^+e^-\to J/\psi +\eta_c]=17.5\pm 5.7$~fb. The quoted
uncertainty reflects only the uncertainties in the values of $m_c$ and
$\langle \bm{p}^2\rangle$. Other theoretical uncertainties are large and
need to be quantified. Nevertheless, our prediction is in agreement with
the Belle and {\it BABAR} results, and it seems that the inclusion of
relative-order-$\alpha_s$ and relativistic corrections resolves the
discrepancy between theory and experiment at the present level of
precision.





\bibliographystyle{aipproc}   


\begin{thebibliography}{9}


\bibitem{Abe:2004ww}
  K.~Abe {\it et al.}  [Belle Collaboration],
  Phys.\ Rev.\ D {\bf 70}, 071102 (2004)
.

\bibitem{Aubert:2005tj}
  B.~Aubert {\it et al.}  [{\it BABAR} Collaboration],
  Phys.\ Rev.\ D {\bf 72}, 031101 (2005)
.

\bibitem{BBL}
G.~T.~Bodwin, E.~Braaten, and G.~P.~Lepage,
{\it Phys.\ Rev.\ } {\bf D51}, 1125 (1995) 
;
\textbf{55}, 5855(E) (1997).

\bibitem{Braaten:2002fi}
E.~Braaten and J.~Lee,
{\it Phys.\ Rev.\ } {\bf D67}, 054007 (2003)
.

\bibitem{Liu:2002wq}
K.~Y.~Liu, Z.~G.~He, and K.~T.~Chao,
{\it Phys.\ Lett.\ } {\bf B557}, 45 (2003)
.

\bibitem{Zhang:2005ch}
  Y.~J.~Zhang, Y.~j.~Gao, and K.~T.~Chao,
  Phys.\ Rev.\ Lett.\  {\bf 96}, 092001 (2006)
.

\bibitem{Gremm:1997dq}
M.~Gremm and A.~Kapustin,                                                
Phys.\ Lett.\ B {\bf 407}, 323 (1997)                                    
. 

\bibitem{Eichten:1978tg}                                              
  E.~Eichten, K.~Gottfried, T.~Kinoshita, K.~D.~Lane, and T.~M.~Yan,   
  Phys.\ Rev.\ D {\bf 17}, 3090 (1978); {\bf 21}, 313(E) (1980).                             

\bibitem{Bodwin:2006dn}
  G.~T.~Bodwin, D.~Kang, and J.~Lee,
  Phys.\ Rev.\ D {\bf 74}, 014014 (2006)
.

\bibitem{Brambilla:1999xf}                                               
  N.~Brambilla, A.~Pineda, J.~Soto, and A.~Vairo,                     
  Nucl.\ Phys.\ B {\bf 566}, 275 (2000)                          
.                                          

\bibitem{Yao:2006px}
  W.~M.~Yao {\it et al.}  [Particle Data Group],
  J.\ Phys.\ G {\bf 33}, 1 (2006).

\end{thebibliography}

%


\end{document}